 \documentclass[aps,preprint,superscriptaddress]{revtex4-1}
  \usepackage[T1]{fontenc}
  \usepackage[english]{babel}
  \usepackage{times}
  \usepackage{amssymb,amsmath,amsfonts,amsthm}
  \usepackage{mathtools}
  \usepackage{verbatim}
  \usepackage{enumerate}
  \usepackage{graphicx}
  \usepackage{hyperref}
  \usepackage{bbm}
  \usepackage{booktabs}
  
  \usepackage[caption=false]{subfig}
  \usepackage{mathrsfs}
  \usepackage{color}
  \usepackage{bm}
  \usepackage{braket}
  \usepackage{euscript}
\usepackage{graphicx}
\usepackage{multirow}
\usepackage{bm}

\usepackage{etoolbox}
\makeatletter
\patchcmd{\frontmatter@abstract@produce}
  {\vskip200\p@\@plus1fil
   \penalty-200\relax
   \vskip-200\p@\@plus-1fil}
  {}
  {}
  {} 
\makeatother
\begin{document}

\title{Quantum Computing with $\mathbb{Z}_{2}$ Abelian anyon system}
\author{Yuanye Zhu}
\thanks{Corresponding author: zhuyy16@mails.tsinghua.edu.cn}
\affiliation{State Key Laboratory of Low-dimensional Quantum Physics, Beijing, 100084, China}
\affiliation{Department of Physics, Tsinghua University, Beijing, 100084, China}

\begin{abstract}
Topological quantum computers provide a fault-tolerant method for performing quantum
computation. Topological quantum computers manipulate topological defects with exotic exchange statistics
called anyons. The simplest anyon model for universal
topological quantum computation is the Fibonacci anyon model, which is a non-abelian anyon system.  In non-abelian anyon systems, exchanging anyons always results a unitary operations instead of a simple phase changing in abelian anyon systems. So, non-abelian anyon systems are of the interest to topological quantum computation. Up till now, most people still  hold the belief that topological quantum computions  can be implemented  only on the non-abelian anyon systems. But actually this is not true. Inspired by extrinsic semiconductor technology, we suggest that abelian anyon systems with defects also support topological quantum computing. In this letter, we report a topological quantum computer prototype based on $\mathbb{Z}_{2}$ abelian anyon system. 

\end{abstract}

\maketitle

\paragraph{Introduction}

\begin{figure}[h]
\centering
\includegraphics[scale=0.3]{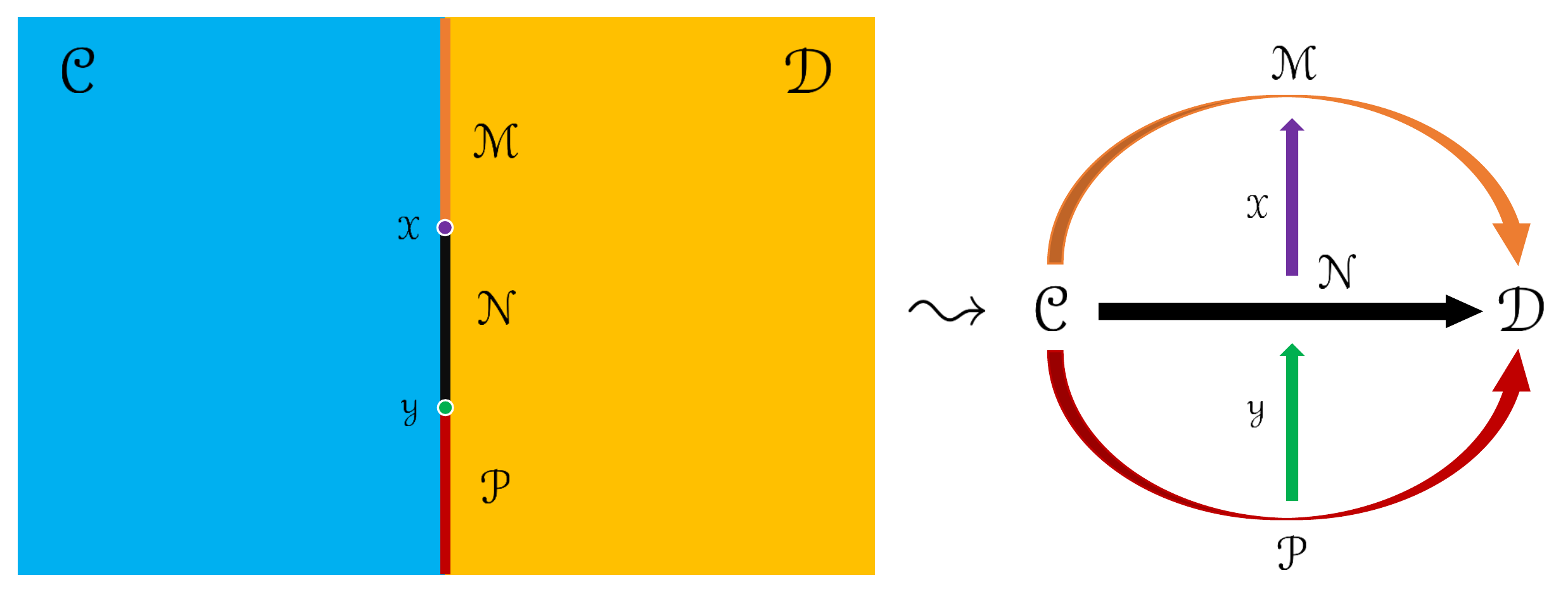}
\caption{2+1D topological order and its category description }\label{2cat}
\end{figure}

In recent years, more and more attention has been paid to the study of topological phase of matter because they transcend Landau's  paradigm of phases and phase transition. Landau's paradigm is based on  symmetry and symmetry breaking theory\cite{Landau1937OnTT}, and its mathematical description is group theory. The new phase of matter requires us to find a new mathematical language that goes beyond group theory to describe the characteristics of topological order. The theory to characterize the topological order is to use the $\mathcal{S}$, $\mathcal{T}$ modular matrices obtained from the non-abelian geometric phases of degenerate ground states on torus and the fusion rules $\mathcal{N}^{ij}_{k}$, such a theory of fusion and braiding of anyons in topological order is so called unitary modular tensor category (UMTC) theory.
For anyon $i$, we use quality $d_i$ to measure the topological degeneracy of $i$, called quantum dimension. An anyon with $d_i=1$ is called abelian anyon, while with $d_i > 1$ is called non-abelian anyon. For a topological order if all the anyon is abelian anyon, this order is called abelian topological order. For a topological order the language of the category can describe all its properties. For example, Figure\ref{2cat}.

The topological order diagram on the left, and its corresponding categorical description on the right. Of these, $\EuScript{C}$ and $\EuScript{D}$ are two UMTC categories representing two different topological orders. $\EuScript{M}$, $\EuScript{N}$, $\EuScript{P}$ are the 1d domain wall between 2d topological order $\EuScript{C}$ and $\EuScript{D}$. In category theory $\EuScript{M} $, $\EuScript{N} $, $\EuScript{P} $ are the functors from category $\EuScript{C}$ to category $\EuScript{D}$. $\EuScript{X}$ is the 0d domain wall between two 1d domain walls  $\EuScript{M}$ and $\EuScript{N}$. In categorical languages, $\EuScript{X}$ is a natural transformation between the functor $\EuScript{M}$ and  $\EuScript{N}$. Further, $\EuScript{M}$, $\EuScript{N}$, $\EuScript{P}$ are all $\EuScript{C}\boxtimes \EuScript{D}^{\mathrm{rev}}$-Unitary fusion left module category $\EuScript{X}$ is $\EuScript{M}\boxtimes_{\EuScript{C}\boxtimes \EuScript{D}^{\mathrm{rev}}} \EuScript{N}^{\mathrm{rev}}$-multi- fusion left module category.

Topological quantum computing cannot be developed without studying topological order\cite{Wen1990TopologicalOI}.  A most remarkable feature of topological phase is that many quantum states are degeneracy in the sense of topology\cite{Wen1990GroundstateDO}(Note: The topological order here refers to topological quantum states with long entanglement based on topological field theory\cite{Chen2010LocalUT}, not topological insulators with only short entanglement \cite{Moore2010TheBO}).
This degeneracy is robust against any local perturbation, so these state gives the possible application as qubits because they are inherently immune to local noise.

 The coding space of topological quantum computation is the  Hilbert space of anyons, which is rather unusual. It is the space of states that corresponds to the fusion process. States that correspond to different fusion outcome are automatically orthogonal to each other as we can always distinguish between different anyons. For abelian topological order, anyons have only one single fusion outcome. So, their fusion Hilbert space is trivial with dimension 1 and the braiding of anyons only comes out a simple phase changing instead of a unitary operator compared with non-abelian topological order. So for abelian topological order the $R$ and $F$ matrix is just the $1 \times 1$ matrix, which is not a operator. So, people hold the belief that topological quantum computation can be implemented only on the non-abelian topological order. However, this is not true. The defects of the abelian topological order efficiently behaves same as composite non-abelian anyons. Up to now, there are still doubts about the existence of non-abelian anyons\cite{Zhang2018QuantizedMC,Zhang2021RetractionNQ}, while the extrinsic topological defects of Abelian order\cite{Wen1991MeanfieldTO,Moroz2017TopologicalOS,Hansson2004SuperconductorsAT,Qi2013AxionTF}  can also support topological degeneracies in 2 + 1D system. So, this gives another possible way to implement the topological quantum computation.

In this letter, we apply our theory to a specific physical example, the $\mathbb{Z}_2$ topological order and give a  gate set, which shows the example of the extra computational power from extrinsic defect of abelian topological order. This new direction not only opens up a new era of theoretical study, but is also instructive for experimental realization of topological quantum computation.

\paragraph{Realization of abelian topological order by bilayer non-chiral fractional quantum hall liquid.} 
The abelian fractional quantum hall states can be classified by $K$ integer valued symmetric matrices, up to an equivalence condition $K \sim O^{T}KO$, where $O \in SL(N,Z)$. All the properties of quasi-particle can be obtained from $K$ matrice, such as statistics and charges. The effective Chern-Simons theory of abelian fractional quantum hall states is 
\begin{equation}
\mathcal{L}=\frac{1}{4\pi}K_{IJ}\epsilon ^{\mu\nu\lambda}a_{I\nu}\partial_{\nu}a_{J\lambda},
\end{equation}
where $a_{I}$ are fictitious $U(1)$ gauge fields.
The $\mathbb{Z}_2$ topological order can be realized in bilayer fractional quantum Hall (FQH) systems. Considers an electron-hole bilayer FQH system, with a $1/2$ Laughlin state of opposite chirality in each layer. The topological order in this system is $U(1)_2\times U(1)_{-2}$, which is equivalent to $\mathbb{Z}_2$ topological order.The superconductor discovered in 1911 was the first experimentally discovered topologically ordered state with $\mathbb{Z}_2$ topological order.

\paragraph{Categorical description of topological order}
At the microscopic level, topological order is the temperature of a quantum system with a zero gap. Macroscopically, a topological sequence can be described by its topological excitations (also known as topological defects). Topological excitation is an excitation that does not create or annihilate from the ground state by local operations.

\begin{figure}[h]
\begin{center}
\includegraphics[scale=0.4]{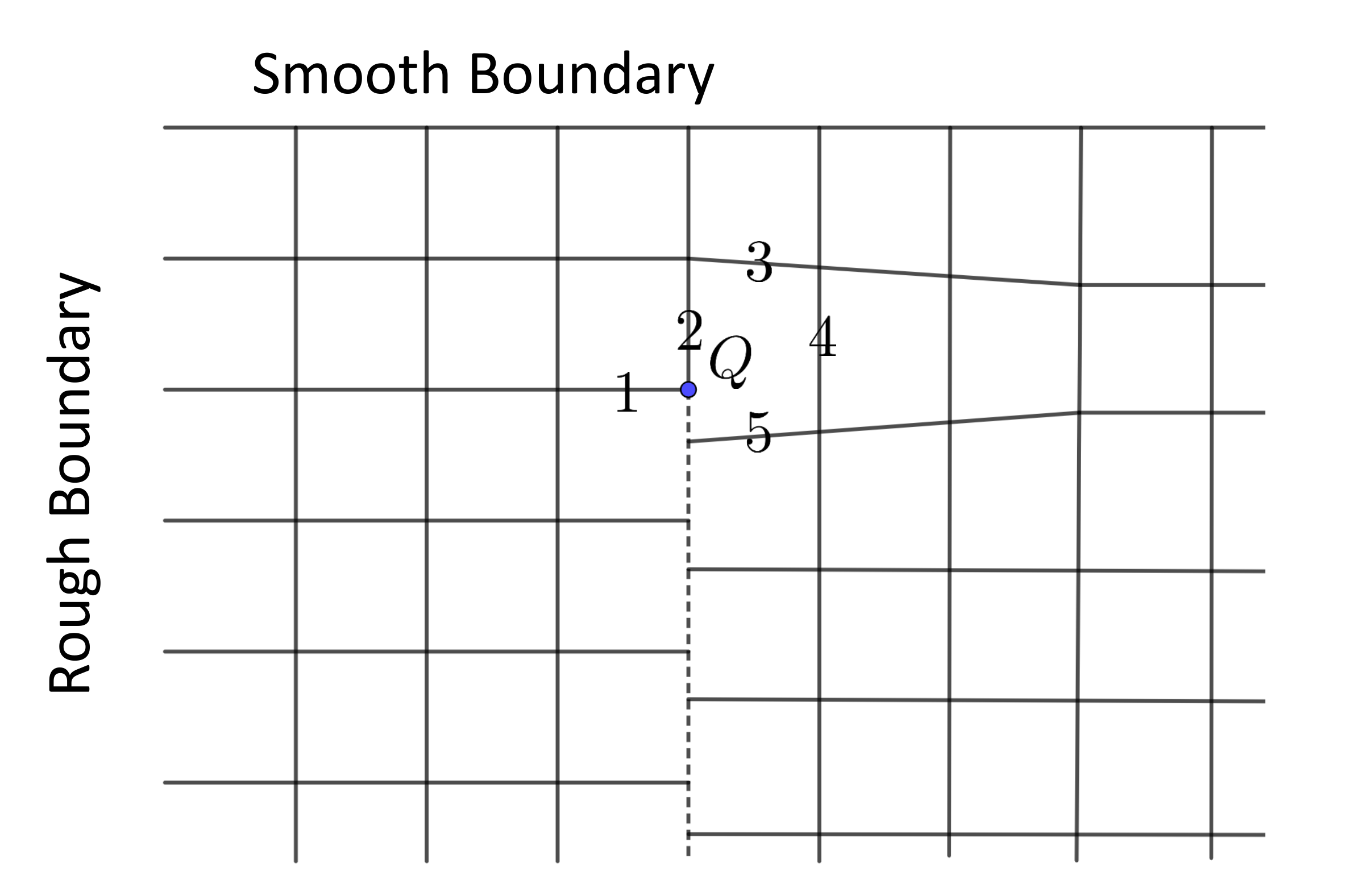}
  \caption{$\mathbb{Z}_{2}$ topological order}\label{ztp}
  \end{center}
  \end{figure}

For $\mathbb{Z}_{2}$ topological order can be described unitary modular tensor category, with following structure:
\begin{itemize}
\item[1.] simple object are 4 anyons $\mathbf{1}, e, m, f$.
\item[2.] fusion rule is $e\otimes m=f$, $e \otimes e=m \otimes m=f \otimes f=\mathbf{1}$
\item[3.] T-matrix
\begin{equation}
\mathcal{T}=\left(\begin{array}{cccc}
1 & 0 & 0 & 0 \\
0 & 1 & 0 & 0 \\
0 & 0 & 1 & 0 \\
0 & 0 & 0 & -1
\end{array}\right)
\end{equation}
\item[4.] S-matrix
\begin{equation}
\mathcal{S}=\frac{1}{2}\left(\begin{array}{cccc}
1 & 1 & 1 & 1 \\
1 & 1 & -1 & -1 \\
1 & -1 & 1 & -1 \\
1 & -1 & -1 & 1
\end{array}\right)
\end{equation}
\end{itemize}
For $\mathbb{Z} _{2}$ topological order, the bulk is described by the category $\mathbf{Toric}$, which has   two condensation algebras, $A_{1} =\mathbf{1}\otimes e$, and $A_{2}=\mathbf{1}\otimes m $. The local $A$ module category of $\mathbf{Toric}$ is $\operatorname{Rep}\left(\mathbb{Z}_{2}\right)$ and $\mathrm{Hilb}_{\mathbb{Z}_{2}}$. According to anyon condensation theory\cite{Kong2014AnyonCA}, this indicates that the $\mathbb{Z}_{2}$ topological order has two boundaries. One is the smooth boundary represented by $\operatorname{Rep}\left(\mathbb{Z}_{2}\right)$ category, whose simple objects are $1,e$, the other is the rough boundary described by $\mathrm{Hilb}_{\mathbb{Z}_{2}}$ category whose simple objects are $1,m$. Corresponding to the figure below Figure\ref{ztp}. You can see $Q$ as a point defect. This can be seen as a 0d domain wall between two 1d domain walls. The stabilizer operator corresponding to this defect is to be commutative with the the stabilizer of toric code Therefore, the operator Q is defined as:  
\begin{equation}
Q=\sigma^{1}_{x}\sigma^{2}_{y}\sigma^{3}_{z}\sigma^{4}_{z}\sigma^{5}_{z}
\end{equation}
This point defect corresponds to anyon $\chi_{\pm}$.
The statistical rule of $\chi_{\pm} $cannot be obtained from this information alone. However, it is well known that the $\mathbb{Z}_{2}$ topological order can be condensed by a double Ising model.Two condensation algebras of double Ising model is $R(\mathbf{1})=\oplus_{i \in \mathrm{Irr}(\mathbf{Ising})}i^{*} \boxtimes i$ and $A=\mathbf{1}\boxtimes \mathbf{1} \oplus \psi \boxtimes \psi$. According to anyon condensation theory, Artical\cite{Chen2020TopologicalPT} gives the structure of double Ising model after condensation, whose objects are:
\begin{equation}
\begin{aligned}
\mathbf{1}:&=A=(\mathbf{1} \otimes \mathbf{1}) \oplus(\psi \otimes \psi)\\
 e:&=(\sigma \otimes \sigma)\\
 m:&=(\sigma \boxtimes \sigma)^{\mathrm{tw}}\\
f:&=(\psi \otimes \mathbf{1}) \otimes A=\psi \otimes \mathbf{1} \oplus \mathbf{1} \otimes \psi\\
\chi_{+}:&=(\mathbf{1} \otimes \sigma) \otimes A=(\mathbf{1} \otimes \sigma) \oplus(\psi \otimes \sigma) \\ \chi_{-}:&=(\sigma \otimes \mathbf{1}) \otimes A=(\sigma \otimes \mathbf{1}) \oplus(\sigma \otimes \psi)\label{611}
\end{aligned}
\end{equation}
The fusion rule for the defect of $\mathbb{Z}_{2}$ topological order can be obtained according to Eq.(\ref{611}):
\begin{equation}
\begin{aligned}
\chi_{\pm} \otimes \chi_{\pm}&=\mathbf{1} \oplus f\\
\chi_{\pm} \otimes \chi_{\mp}&=e \oplus m\\
 e \otimes \chi_{\pm}&=\chi_{\pm} \otimes e=m \otimes \chi_{\pm}=\chi_{\pm} \otimes m=\chi_{\mp}
\end{aligned}
\end{equation}

\paragraph{Topological quantum computation with defects of $\mathbb{Z}_{2}$ topological order} 
To initialize a quantum computer, one needs first to identify the computational space of n-qubits. In topological computer this means creating some number of anyons from the vacuum and fixing their positions. The system then exhibits a fusion space manifesting as a protected non-local subspace that is identified as the computational space.Let us focus on a system of six $\chi_{\pm}$-anyons that enables to demonstrate all the basic operations. It is convenient to choose the pairwise fusion basis as a computational basis:

\begin{figure}[h]
\centering
\includegraphics[width=0.6\linewidth]{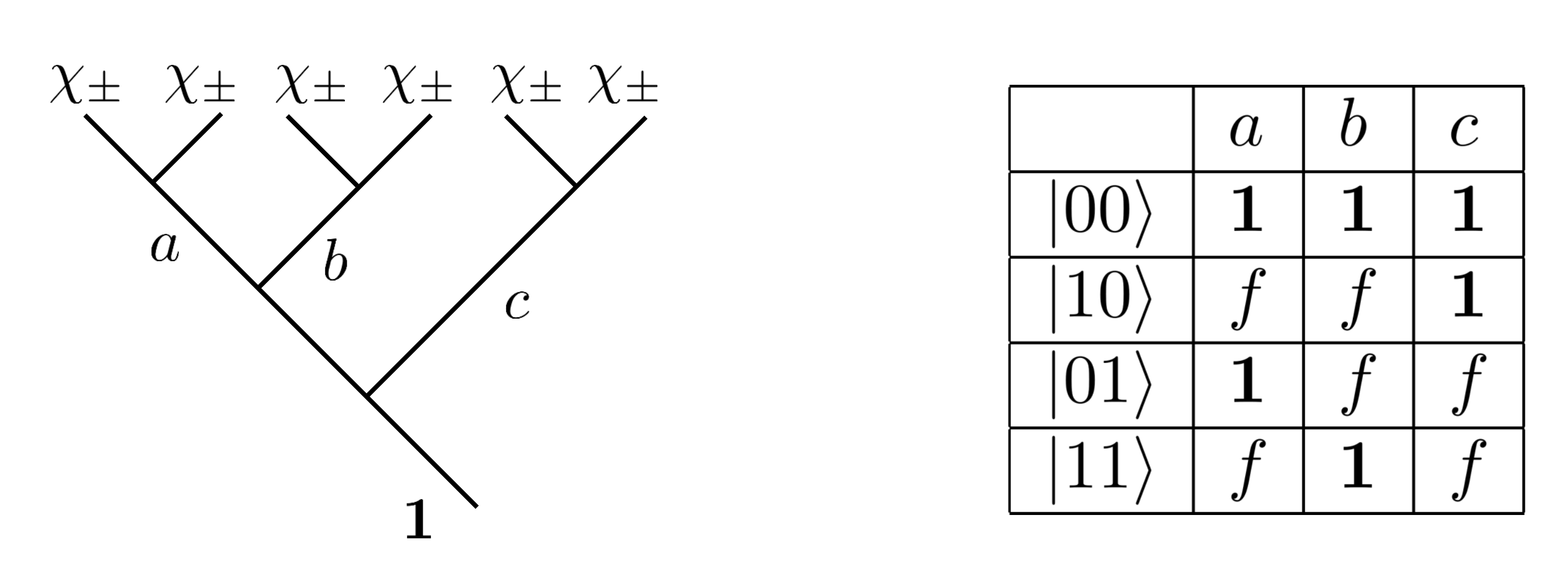}
\caption{Anyon coding}\label{tqcbm}
\end{figure}

Define $F$ and $R$ matrix as follow
\begin{figure}[h]
\centering
\includegraphics[width=0.8\linewidth]{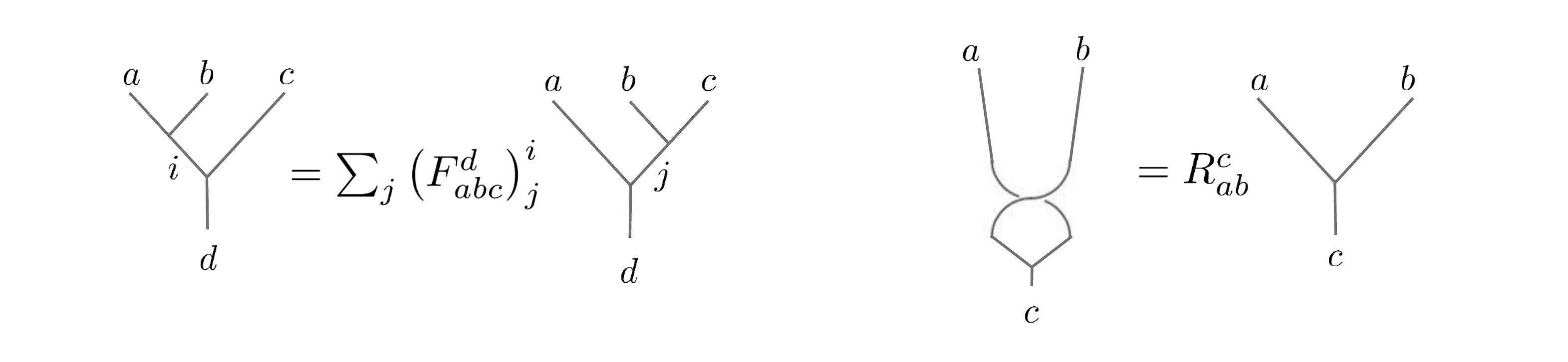}
\caption{$F$-matrix and $R$-matrix}\label{fr}
\end{figure}

\begin{figure}[h]
\centering
\includegraphics[width=0.6\linewidth]{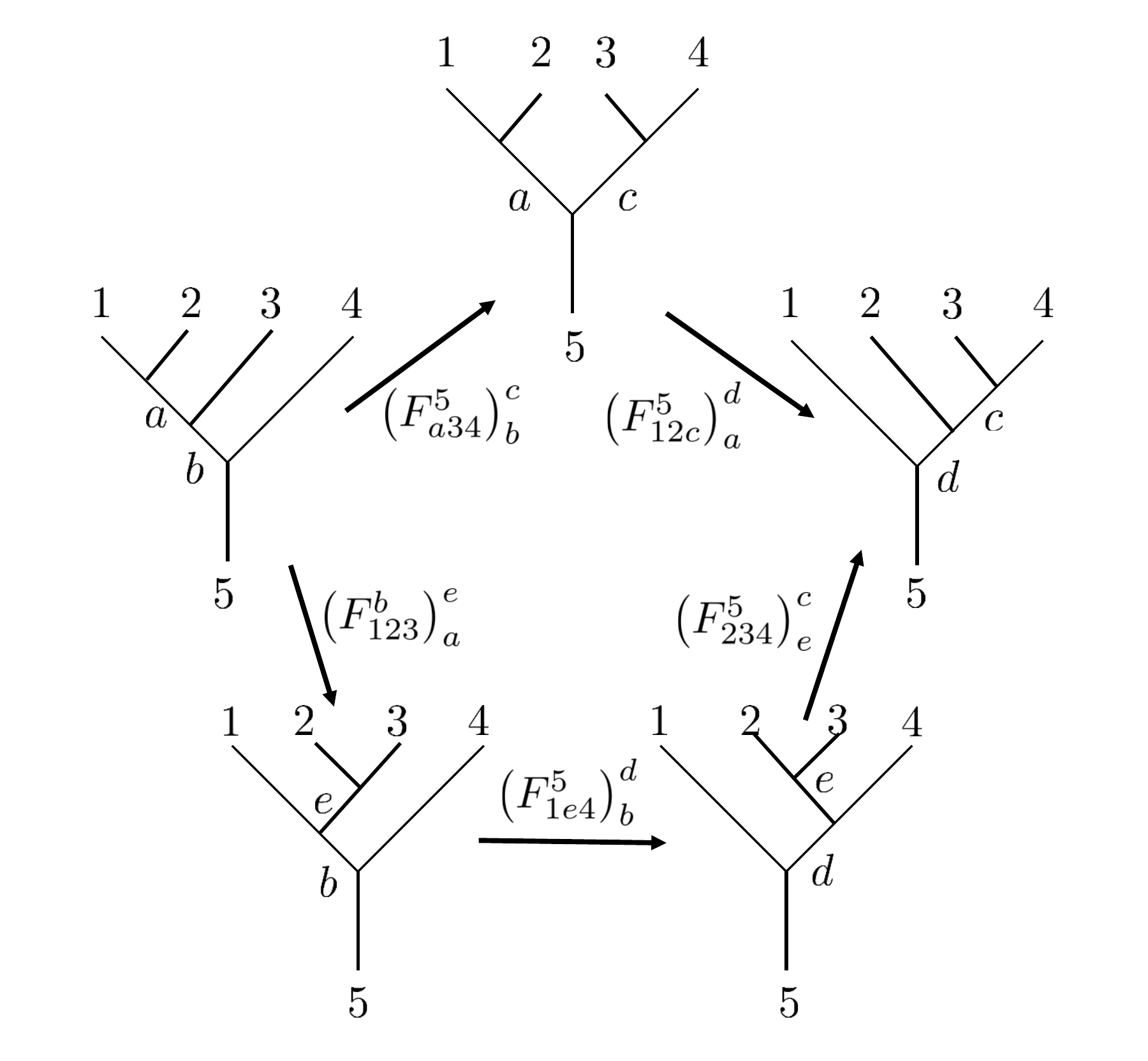}
\caption{Pentagon relation}\label{wbx}
\end{figure}

According to the pentagram relationship (Figure\ref{wbx}.), the following identities are given:
\begin{equation}
\left(F_{12 c}^{5}\right)_{a}^{d}\left(F_{a 34}^{5}\right)_{b}^{c}=\sum_{e}\left(F_{234}^{d}\right)_{e}^{c}\left(F_{1 e 4}^{5}\right)_{b}^{d}\left(F_{123}^{b}\right)_{a}^{e}\label{wbxeq}
\end{equation}
Here let 1,2,3,4 be point defects $\chi_{\pm}$ and 5 be $\mathbf{1}$, according to the fusion rule of $\mathbb{Z}_{2}$ topological order $b$ and $d$ must be $\chi_{pm}$, $a$ and $c$ can be $\mathbf{1}$ or $f$.
When $a=c =\mathbf{1}$, the formula Fq.(\ref{wbxeq}) can be written:
\begin{equation}
\left(F_{\chi_{\pm}, \chi_{\pm}, \mathbf{1}}^{\mathbf{1}}\right)_{\mathbf{1}}^{\chi_{\pm}}\left(F_{\mathbf{1}, \chi_{\pm}, \chi_{\pm}}^{\mathbf{1}}\right)_{\chi_{\pm}}^{\mathbf{1}}=\sum_{e=\mathbf{1}, f}\left(F_{\chi_{\pm},\chi_{\pm},\chi_{\pm}}^{\chi_{\pm}}\right)_{e}^{\mathbf{1}}\left(F_{\chi_{\pm} ,e ,\chi_{\pm}}^{\mathbf{1}}\right)_{\chi_{\pm}}^{\chi_{\pm}}\left(F_{\chi_{\pm},\chi_{\pm},\chi_{\pm}}^{\chi_{\pm}}\right)_{\mathbf{1}}^{e}\label{608}
\end{equation}
where
$\left(F_{\chi_{\pm}, \chi_{\pm}, \mathbf{1}}^{\mathbf{1}}\right)_{\mathbf{1}}^{\chi_{\pm}}$,$\left(F_{\mathbf{1}, \chi_{\pm}, \chi_{\pm}}^{\mathbf{1}}\right)_{\mathbf{1}}^{\chi_{\pm}}$,$\left(F_{\chi_{\pm},\mathbf{1}, \chi_{\pm}}^{\mathbf{1}}\right)_{\chi_{\pm}}^{\chi_{\pm}}$ are the rearrangement of anyons, which are trival, so these functors correspond to identity naturally. 

$\left(F_{\chi_{\pm}, f,\chi_{\pm}}^{\mathbf{1}}\right)_{\chi_{\pm}}^{\chi_{\pm}}$ is clearly a identity functor. Therefore, so that all of the above functions are mapped in units, the formula Eq.(~\ref{608}) can be reduced to:
\begin{equation}
1={\left(F_{\chi_{\pm}, \chi_{\pm}, \chi_{\pm}}^{\chi_{\pm}}\right)_{\mathbf{1}}^{\mathbf{1}}}^{2}+\left(F_{\chi_{\pm}, \chi_{\pm}, \chi_{\pm}}^{\chi_{\pm}}\right)_{f}^{\mathbf{1}}\left(F_{\chi_{\pm}, \chi_{\pm}, \chi_{\pm}}^{\chi_{\pm}}\right)_{\mathbf{1}}^{f}\label{609}
\end{equation}
When $a=\mathbf{1}$, $c=f$, Eq.(\ref{wbxeq}) can be written as:
\begin{equation}
\left(F_{\chi_{\pm}, \chi_{\pm}, f}^{\mathbf{1}}\right)_{\mathbf{1}}^{\chi_{\pm}}\left(F_{\mathbf{1}, \chi_{\pm}, \chi_{\pm}}^{f}\right)_{\chi_{\pm}}^{\mathbf{1}}=\sum_{e=\mathbf{1}, f}\left(F_{\chi_{\pm},\chi_{\pm},\chi_{\pm}}^{\chi_{\pm}}\right)_{e}^{f}\left(F_{\chi_{\pm} ,e ,\chi_{\pm}}^{\mathbf{1}}\right)_{\chi_{\pm}}^{\chi_{\pm}}\left(F_{\chi_{\pm},\chi_{\pm},\chi_{\pm}}^{\chi_{\pm}}\right)_{\mathbf{1}}^{e}\label{610}
\end{equation}
Where $\left(F_{\chi_{\pm}, \chi_{\pm}, f}^{\mathbf{1}}\right)_{\mathbf{1}}^{\chi_{\pm}}$ not obey the fusion rule,it is illegal functor. So, $\left(F_{\chi_{\pm}, \chi_{\pm}, f}^{\mathbf{1}}\right)_{\mathbf{1}}^{\chi_{\pm}}=0$, where $\left(F_{\chi_{\pm}, \chi_{\pm},  \chi_{\pm}}^{ \chi_{\pm}}\right)_{\mathbf{1}}^{f}$ is the identity functor naturally, so $\left(F_{\chi_{\pm}, \chi_{\pm},  \chi_{\pm}}^{ \chi_{\pm}}\right)_{\mathbf{1}}^{f}=1$, and Eq.(\ref{610}) can be simplify as
\begin{equation}
\left(F_{\chi_{\pm}, \chi_{\pm},  \chi_{\pm}}^{ \chi_{\pm}}\right)_{f}^{f}=-\left(F_{\chi_{\pm}, \chi_{\pm},  \chi_{\pm}}^{ \chi_{\pm}}\right)_{\mathbf{1}}^{\mathbf{1}}\label{611}
\end{equation}

When $a=c=f$, Figure\ref{wbxeq}. can be written as
\begin{equation}
\left(F_{\chi_{\pm}, \chi_{\pm}, f}^{\mathbf{1}}\right)_{f}^{\chi_{\pm}}\left(F_{f, \chi_{\pm}, \chi_{\pm}}^{\mathbf{1}}\right)_{\chi_{\pm}}^{f}=\sum_{e=\mathbf{1}, f}\left(F_{\chi_{\pm},\chi_{\pm},\chi_{\pm}}^{\chi_{\pm}}\right)_{e}^{f}\left(F_{\chi_{\pm} ,e ,\chi_{\pm}}^{\mathbf{1}}\right)_{\chi_{\pm}}^{\chi_{\pm}}\left(F_{\chi_{\pm},\chi_{\pm},\chi_{\pm}}^{\chi_{\pm}}\right)_{f}^{e}\label{612}
\end{equation}
Simplified available:
\begin{equation}
1=\left(F_{\chi_{\pm}, \chi_{\pm}, \chi_{\pm}}^{\chi_{\pm}}\right)^{f}_{\mathbf{1}}\left(F_{\chi_{\pm}, \chi_{\pm}, \chi_{\pm}}\right)^{\mathbf{1}}_{f}+{\left(F_{\chi_{\pm}, \chi_{\pm}, \chi_{\pm}}^{\chi_{\pm}}\right)_{f}^{f}}^{2}
\end{equation}

Next, let 1 be $f$; 2,3,4,5 be $\chi_{\pm}$, according to fusion rule $a=d=\chi_{\pm}$,  anyon $b$ and $c$ can be $1$ or $f$. Let $b = c = \mathbf{1}$, pentagon relation Eq.(\ref{wbxeq}) can be written as:
\begin{equation}
\left(F_{f,\chi_{\pm},\mathbf{1}}^{\chi_{\pm}}\right)_{\chi_{\pm}}^{\chi_{\pm}}\left(F_{\chi_{\pm},\chi_{\pm},\chi_{\pm}}^{\chi_{\pm}}\right)_{\mathbf{1}}^{\mathbf{1}}=\sum_{e=\mathbf{1}, f}\left(F_{\chi_{\pm},\chi_{\pm},\chi_{\pm}}^{\chi_{\pm}}\right)_{e}^{\mathbf{1}}\left(F_{f,e, \chi_{\pm}}^{\chi_{\pm}}\right)_{\mathbf{1}}^{\chi_{\pm}}\left(F_{f,\chi_{\pm},\chi_{\pm}}^{\mathbf{1}}\right)_{\chi_{\pm}}^{e}\label{614}
\end{equation}

Where$\left(F_{f,\chi_{\pm},\mathbf{1}}^{\chi_{\pm}}\right)_{\chi_{\pm}}^{\chi_{\pm}}=1$, this is the fusion of two anyons, the rearrangement of which is trical. $\left(F_{f,\chi_{\pm},\chi_{\pm}}^{\mathbf{1}}\right)_{\chi_{\pm}}^{\mathbf{1}}=0$ not obey fusion rule. let $A=\left(F_{f,f,\chi_{\pm}}^{\chi_{\pm}}\right)_{\mathbf{1}}^{\chi_{\pm}}\left(F_{f,\chi_{\pm},\chi_{\pm}}^{\mathbf{1}}\right)_{\chi_{\pm}}^{f}$, Eq.(\ref{614})can be simplification as
\begin{equation}
\left(F_{\chi_{\pm},\chi_{\pm},\chi_{\pm}}^{\chi_{\pm}}\right)_{\mathbf{1}}^{\mathbf{1}}=A\left(F_{\chi_{\pm},\chi_{\pm},\chi_{\pm}}^{\chi_{\pm}}\right)_{f}^{\mathbf{1}}\label{615}
\end{equation}

Finally, let$b=\mathbf{1}, c=f$, pentagon relation Eq.(\ref{wbxeq}) can be written as
\begin{equation}
\left(F_{f,\chi_{\pm},f}^{\chi_{\pm}}\right)_{\chi_{\pm}}^{\chi_{\pm}}\left(F_{\chi_{\pm},\chi_{\pm},\chi_{\pm}}^{f}\right)_{\mathbf{1}}^{\mathbf{1}}=\sum_{e=\mathbf{1}, f}\left(F_{\chi_{\pm},\chi_{\pm},\chi_{\pm}}^{\chi_{\pm}}\right)_{e}^{f}\left(F_{f,e, \chi_{\pm}}^{\chi_{\pm}}\right)_{\mathbf{1}}^{\chi_{\pm}}\left(F_{f,\chi_{\pm},\chi_{\pm}}^{\mathbf{1}}\right)_{\chi_{\pm}}^{e}\label{616}
\end{equation}
where $\left(F_{f,\chi_{\pm},f}^{\chi_{\pm}}\right)_{\chi_{\pm}}^{\chi_{\pm}}=1$, due to the symmetry of fusion, so its rearrangement is trival. Eq.(\ref{616}) can be write as
\begin{equation}
\left(F_{\chi_{\pm},\chi_{\pm},\chi_{\pm}}^{\chi_{\pm}}\right)_{\mathbf{1}}^{f}=A\left(F_{\chi_{\pm},\chi_{\pm},\chi_{\pm}}^{\chi_{\pm}}\right)_{f}^{f}\label{617}
\end{equation}
From Eq.(\ref{609}), Eq.(\ref{611}), Eq.(\ref{615}), Eq.(\ref{617}), $F_{\chi_{\pm},\chi_{\pm},\chi_{\pm}}^{\chi_{\pm}}$and the unitary of matrix, one will get the $F$-matrix (up to global phase):
\begin{equation}
F_{\chi_{\pm},\chi_{\pm},\chi_{\pm}}^{\chi_{\pm}}= \frac{1}{\sqrt{2}}\left(\begin{array}{cc}
1 & 1 \\
1 & -1
\end{array}\right)
\end{equation}

\begin{figure}
\centering
\includegraphics[width=0.8\linewidth]{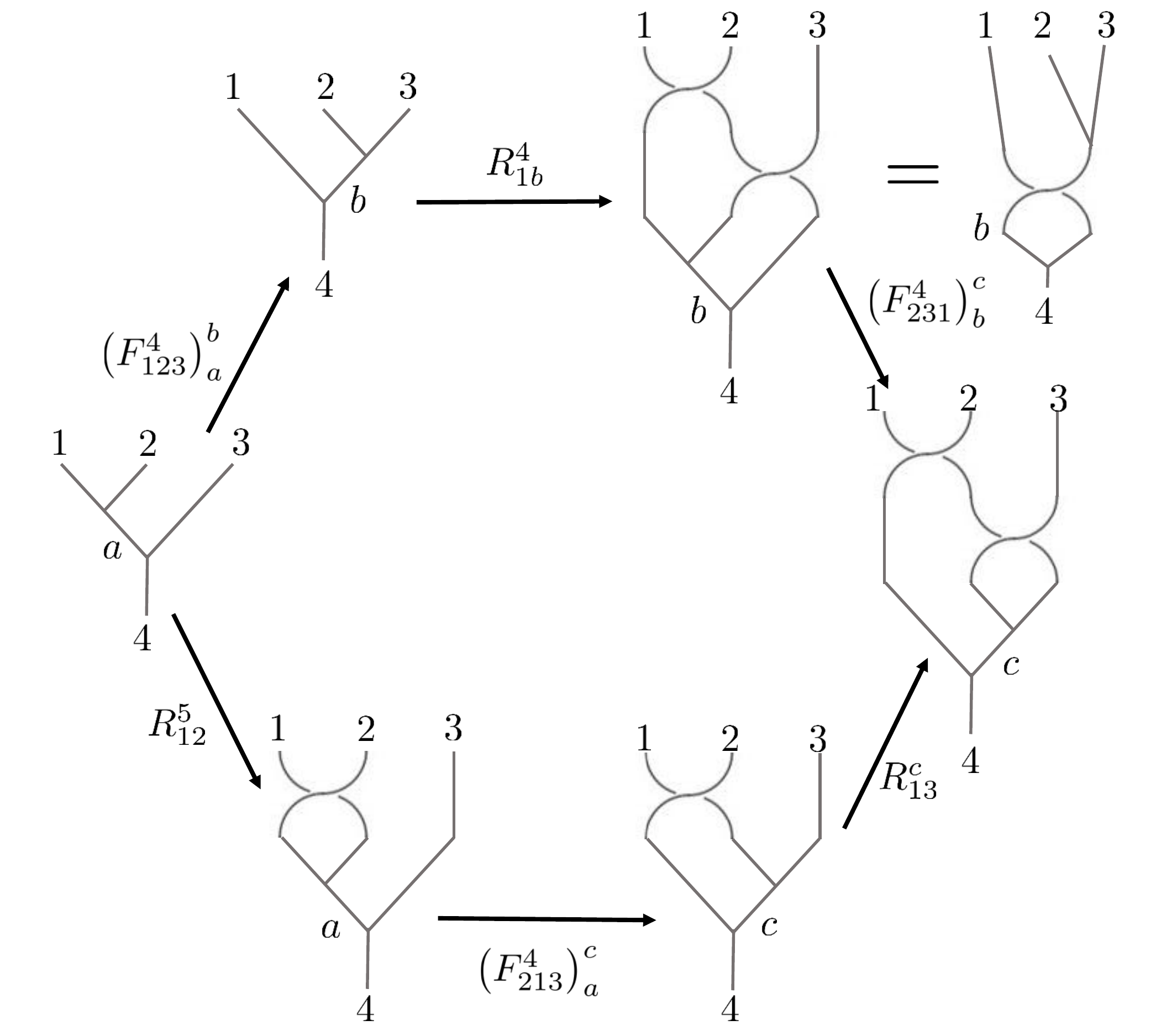}
\caption{Hexagon relation}\label{bbx}
\end{figure}

Next,$R$-matrix will be calculated, Figure\ref{bbx}. is the hexagon relation
which correspondence to the hexagon equation:
\begin{equation}
\sum_{b}\left(F_{231}^{4}\right)_{b}^{c} R_{1 b}^{4}\left(F_{123}^{4}\right)_{a}^{b}=R_{13}^{c}\left(F_{213}^{4}\right)_{a}^{c} R_{12}^{a}\label{bbxx}
\end{equation}
Let 1,2,3,4 be $\chi_{\pm}$, according to fusion rul, $a$ and $c$ can be $\mathbf{1},f$.
When $a=c=\mathbf{1}$,
\begin{equation}
\sum_{b=\mathbf{1},f}\left(F_{\chi_{\pm},\chi_{\pm},\chi_{\pm}}^{\chi_{\pm}}\right)_{b}^{\mathbf{1}} R_{\chi_{\pm} b}^{\chi_{\pm}}\left(F_{\chi_{\pm},\chi_{\pm},\chi_{\pm}}^{\chi_{\pm}}\right)_{\mathbf{1}}^{b}=R_{\chi_{\pm},\chi_{\pm}}^{\mathbf{1}}\left(F_{\chi_{\pm},\chi_{\pm},\chi_{\pm}}^{\chi_{\pm}}\right)_{\mathbf{1}}^{\mathbf{1}} R_{\chi_{\pm}\chi_{\pm}}^{\mathbf{1}}
\end{equation}
made simplification
\begin{equation}
\frac{1}{\sqrt{2}}\left(R_{\chi_{\pm}}^{\chi_{\pm}}+R_{\chi_{\pm} f}^{\chi_{\pm}}\right)=R_{\chi_{\pm} \chi_{\pm}}^{1}{ }^{2}
\end{equation} \label{621}
When $a=c=f$, Eq.(\ref{bbxx}) can be written as
\begin{equation}
\sum_{b=\mathbf{1},f}\left(F_{\chi_{\pm},\chi_{\pm},\chi_{\pm}}^{\chi_{\pm}}\right)_{b}^{f} R_{\chi_{\pm} b}^{\chi_{\pm}}\left(F_{\chi_{\pm},\chi_{\pm},\chi_{\pm}}^{\chi_{\pm}}\right)_{f}^{b}=R_{\chi_{\pm},\chi_{\pm}}^{f}\left(F_{\chi_{\pm},\chi_{\pm},\chi_{\pm}}^{\chi_{\pm}}\right)_{f}^{f} R_{\chi_{\pm}\chi_{\pm}}^{f}
\end{equation}
Simplify to:
\begin{equation}
\frac{1}{\sqrt{2}}\left(R^{\chi_{\pm}}_{\chi_{\pm}\mathbf{1}}+R^{\chi_{\pm}}_{\chi_{\pm}f}\right)=-{R^{f}_{\chi_{\pm}\chi_{\pm}}}^{2}\label{623}
\end{equation}

When $a=\mathbf{1}, c=f$, Eq.(\ref{bbxx}) is
\begin{equation}
\sum_{b=\mathbf{1},f}\left(F_{\chi_{\pm},\chi_{\pm},\chi_{\pm}}^{\chi_{\pm}}\right)_{b}^{f} R_{\chi_{\pm} b}^{\chi_{\pm}}\left(F_{\chi_{\pm},\chi_{\pm},\chi_{\pm}}^{\chi_{\pm}}\right)_{\mathbf{1}}^{b}=R_{\chi_{\pm},\chi_{\pm}}^{f}\left(F_{\chi_{\pm},\chi_{\pm},\chi_{\pm}}^{\chi_{\pm}}\right)_{\mathbf{1}}^{f} R_{\chi_{\pm}\chi_{\pm}}^{\mathbf{1}}
\end{equation}
Simplify to:
\begin{equation}
\frac{1}{\sqrt{2}}\left(R^{\chi_{\pm}}_{\chi_{\pm}\mathbf{1}}-R^{\chi_{\pm}}_{\chi_{\pm}f}\right)=R^{f}_{\chi_{\pm}\chi_{\pm}}R^{\mathbf{1}}_{\chi_{\pm}\chi_{\pm}}\label{625}
\end{equation}
When $a=f, c=\mathbf{1}$, Eq. (\ref{bbxx}) can bewritten as:
\begin{equation}
\sum_{b=\mathbf{1},f}\left(F_{\chi_{\pm},\chi_{\pm},\chi_{\pm}}^{\chi_{\pm}}\right)_{b}^{\mathbf{1}} R_{\chi_{\pm} b}^{\chi_{\pm}}\left(F_{\chi_{\pm},\chi_{\pm},\chi_{\pm}}^{\chi_{\pm}}\right)_{f}^{b}=R_{\chi_{\pm},\chi_{\pm}}^{\mathbf{1}}\left(F_{\chi_{\pm},\chi_{\pm},\chi_{\pm}}^{\chi_{\pm}}\right)_{\mathbf{1}}^{f} R_{\chi_{\pm}\chi_{\pm}}^{f}
\end{equation}
Simplify to:
\begin{equation}
\frac{1}{\sqrt{2}}\left(R^{\chi_{\pm}}_{\chi_{\pm}\mathbf{1}}-R^{\chi_{\pm}}_{\chi_{\pm}f}\right)=R^{\mathbf{1}}_{\chi_{\pm}\chi_{\pm}}R^{f}_{\chi_{\pm}\chi_{\pm}}\label{627}
\end{equation}
where $R^{\chi_{\pm}}_{\chi_{\pm}\mathbf{1}}=1$,because it is the braised with vacuum, which is trivial. From Eq.(\ref{621}), Eq.(\ref{623}), Eq.(\ref{625}), Eq.(\ref{627}), $R$-matrix will get:
\begin{equation}
R_{\chi_{\pm}\chi_{\pm}}=e^{-i \pi / 8}\left(\begin{array}{cc}1 & 0 \\ 0 & i\end{array}\right)
\end{equation}

If anyon $a$ and $b$don't have direct fusion channel, their braise can be defined as follow
\begin{figure}[h]
\centering
\includegraphics[width=0.8\linewidth]{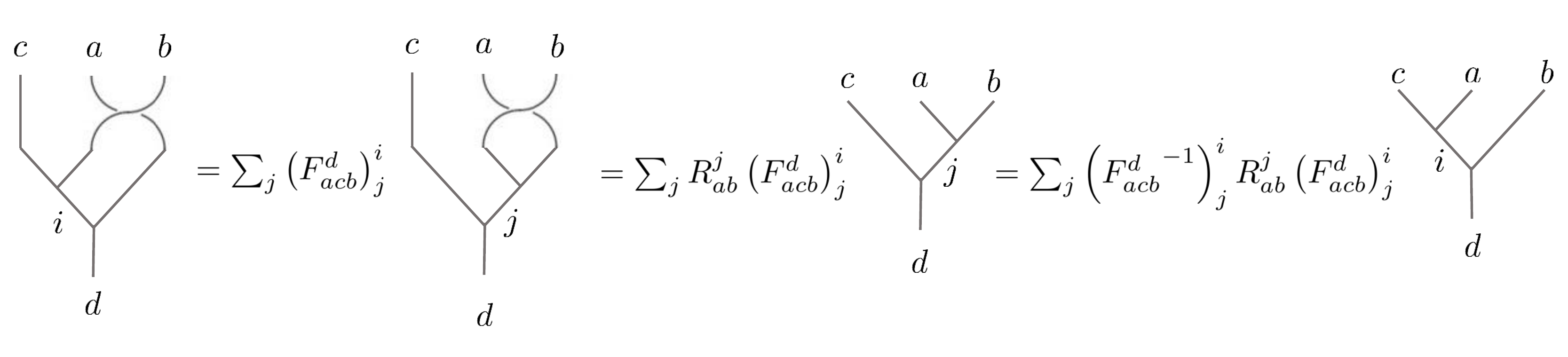}
\caption{$B$-matrix}\label{bmm}
\end{figure}
The graph above shows the $B $-matrix:
\begin{equation}
B=F^{-1}RF
\end{equation}

\begin{figure}[h]
\includegraphics[width=\linewidth]{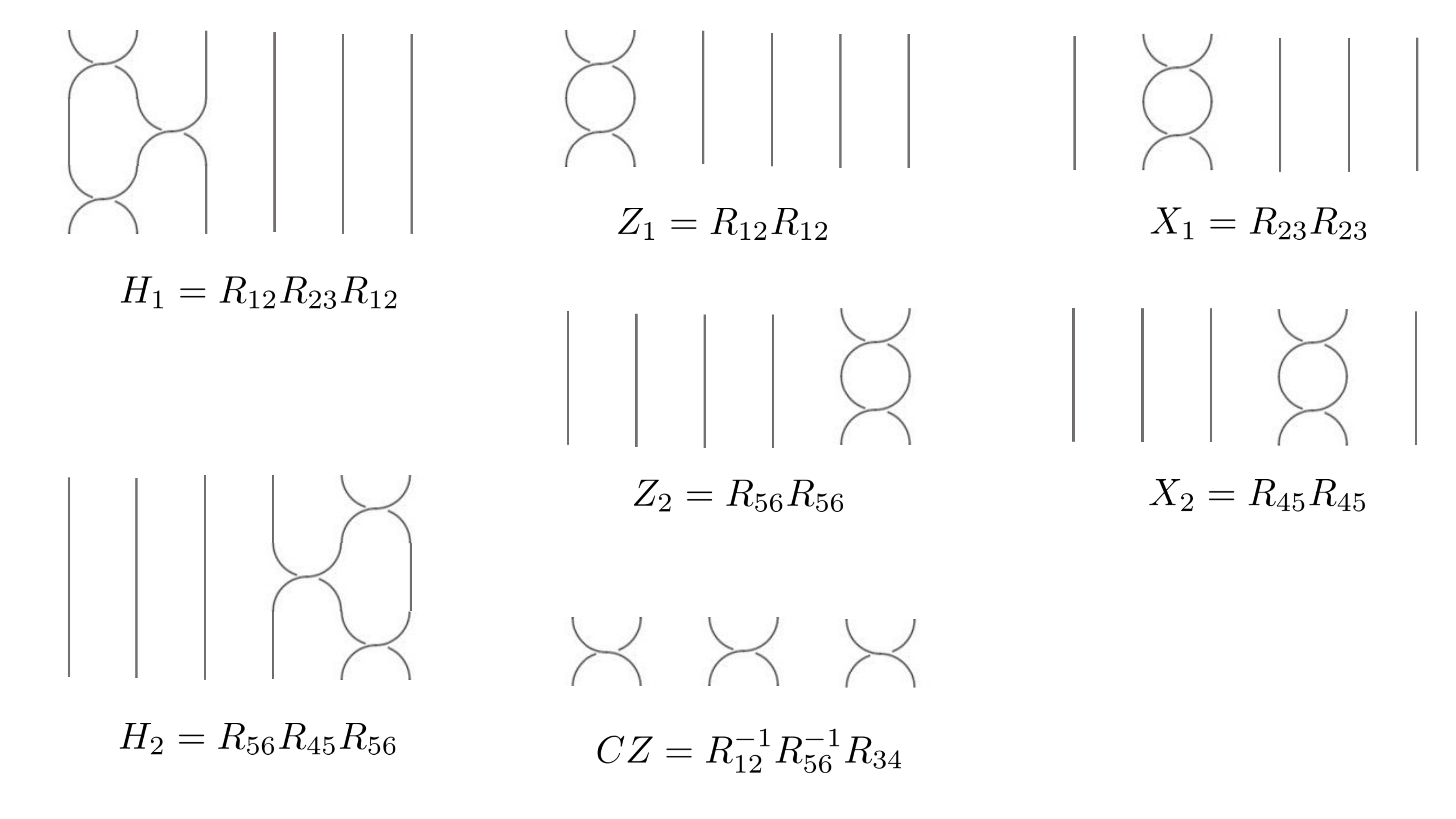}
\caption{Manipulate of quantum gate in topological quantum computer}\label{bzq}
\end{figure}

Thus, common single-qubit gate  can be constructed with a matrix of $R$, $F$, and $B$ matrices (up to global  phase), with the following results:
\begin{equation}
\begin{array}{lll}
X_{1}= & R_{23}^{2}=F^{-1} R^{2} F \otimes I, & Z_{1}=R_{12}^{2}=R^{2} \otimes I \\
X_{2}= & R_{45}^{2}=I \otimes F^{-1} R^{2} F, & Z_{2}=R_{56}^{2}=I \otimes R^{2}
\end{array}
\end{equation}
\begin{equation}
\begin{array}{l}
U_{H, 1}=R_{12} R_{23} R_{12}=R F^{-1} R F R \otimes I \\
U_{H, 2}=R_{56} R_{45} R_{56}=I \otimes R F^{-1} R F R
\end{array}
\end{equation}
\begin{equation}
U_{C Z}=R_{12}^{-1} R_{34} R_{56}^{-1}
\end{equation}
For these  quantum gates, the corresponding braiding operations are shown in Figure\ref{bzq}.

\begin{figure}[h]
\includegraphics[width=\linewidth]{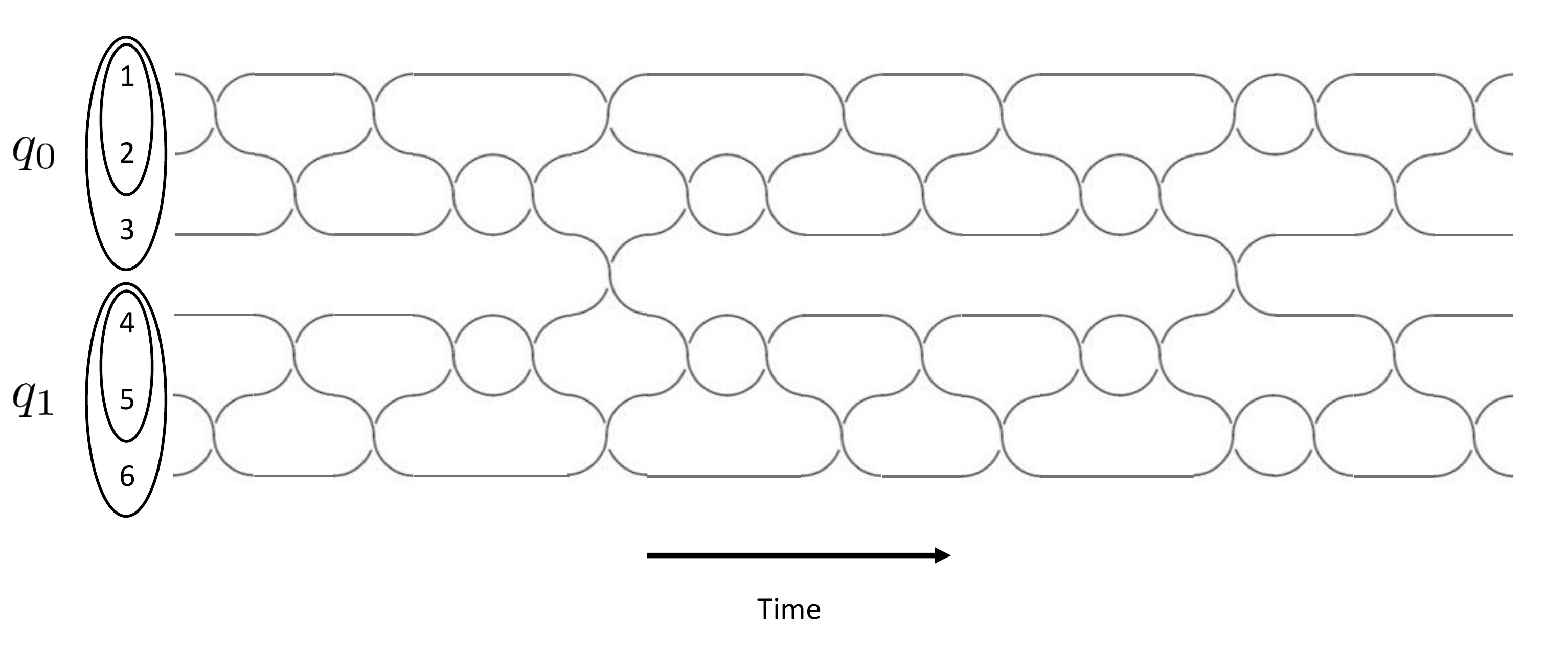}
\caption{Topological quantum computation of Grover algorithm}\label{tqcga}
\end{figure}

Take example of Grover algorithm with target state $|w\rangle=|00\rangle$, below is the quantum circuit
\begin{figure}[h]
\includegraphics[width=\linewidth]{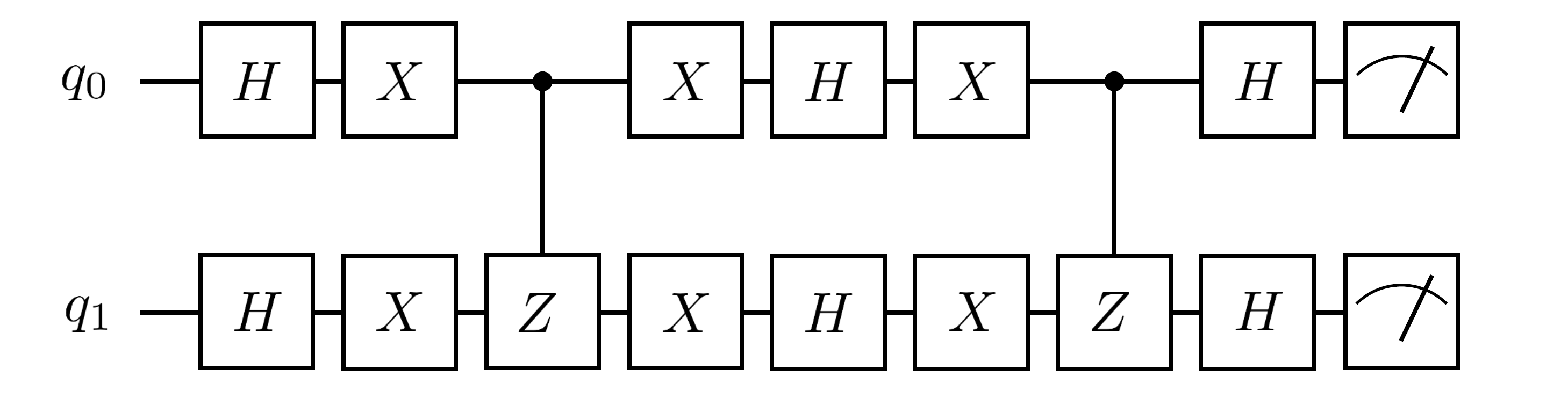}
\caption{Quantum circuit of Grover algorithm}\label{2GL}
\end{figure}
This circuit  can be implemented by topological quantum computing proposed in this article, and the braiding rules are as follows Figure\ref{tqcga}.

\paragraph{Conclusions} Abelian topological order with  external defects supports a  topological quantum computing.  In this Letter, we  take  $\mathbb{Z}_2$ topological order as a concrete physical example. $\mathbb{Z}_2$ topological order can be realized by bilayer fractional quantum Hall liquids and superconductor systems.  The challenge of realistically implementing our  universal quantum gate now lies in how to execute the realistic control.

\bibliographystyle{apsrev4-1}
\bibliography{refs}
\end{document}